\documentclass[final,5p,times,twocolumn]{elsarticle}

\usepackage{amssymb}
\usepackage{amsmath}

\biboptions{sort&compress}

\usepackage{threeparttable}
\usepackage{xcolor}

\journal{Nuclear Instruments and Methods in Physics Research Section B: Beam Interactions with Materials and Atoms}

\begin{document}

\begin{frontmatter}



\title{Measurement of radionuclide production probabilities in negative muon nuclear capture and validation of Monte Carlo simulation model}


\author[jaea]{Y.~Yamaguchi}
\author[riken]{M.~Niikura}
\author[ut]{R.~Mizuno}
\author[kek,kekj]{M.~Tampo}
\author[jaea]{M.~Harada}
\author[kek,kekj]{N.~Kawamura}
\author[kek,kekj]{I.~Umegaki}
\author[kek,kekj]{S.~Takeshita}
\author[jaea]{K.~Haga}

\affiliation[jaea]{organization={J-PARC Center, Japan Atomic Energy Agency},
            addressline={2-4 Shirakata}, 
            city={Tokai},
            postcode={319-1195}, 
            state={Ibaraki},
            country={Japan}}

\affiliation[riken]{organization={RIKEN Nishina Center},
            addressline={2-1 Hirosawa}, 
            city={Wako},
            postcode={351-0198}, 
            state={Saitama},
            country={Japan}}

\affiliation[ut]{organization={Department of Physics, The University of Tokyo},
            addressline={7-3-1 Hongo}, 
            city={Bunkyo},
            postcode={113-0033}, 
            state={Tokyo},
            country={Japan}}

\affiliation[kek]{organization={Institute of Materials Structure Science, High Energy Accelerator Research Organization (KEK)},
            addressline={1-1 Oho}, 
            city={Tsukuba},
            postcode={305-0801}, 
            state={Ibaraki},
            country={Japan}}

\affiliation[kekj]{organization={J-PARC Center, High Energy Accelerator Research Organization (KEK)},
            addressline={2-4 Shirakata}, 
            city={Tokai},
            postcode={319-1195}, 
            state={Ibaraki},
            country={Japan}}
\begin{abstract}
As part of the development of a sample radioactivity calculation program, we have measured radionuclide production probabilities in negative muon nuclear capture to update experimental data and to validate a calculation dataset obtained by a Monte Carlo simulation code. 
The probabilities have been obtained by an activation experiment on $^{27}$Al, $^\mathrm{nat}$Si, $^{59}$Co, and $^\mathrm{nat}$Ta targets.
The obtained probabilities expand the validation scope to the radionuclide production processes outside of the existing data coverage.
By comparing the resultant probabilities with the calculated dataset, it has been revealed that the dataset is generally on the {\color{black}safe} side in radioactivity estimation and needs to be corrected in the following three cases: (i) isomer production; (ii) radionuclide production by the multiple neutron emission; (iii) radionuclide production by particle emissions involving a proton.
The present probabilities and the new findings on the correction provide valuable clues to improvements of the simulation models.
\end{abstract}




\end{frontmatter}



\section{Introduction}
\label{sec:intro}

Negative muon nuclear capture is one of the {\color{black}reactions to consider} in high-intensity and high-energy accelerator facility design and operation~\cite{SAKAKI2020164323, Grillenberger, jparc_1999} due to radionuclide production.
The radionuclide is produced by the negative muon ($\mu^-$) generated in the accelerator facilities through the following processes.
The $\mu^-$ stopped in matter is captured into an atomic orbit of a nucleus and cascades down to the 1s state of the muonic atom by emitting Auger electrons and muonic X-rays.
From the 1s state of the muonic atom, the $\mu^-$ either decays into an electron and two neutrinos or is captured by the nucleus, which is called negative muon nuclear capture.
The $\mu^-$ nuclear capture reaction is the major process for the nucleus with atomic number $Z > 12$.
This reaction produces a highly excited nucleus with $(Z-1)$ because a part of the $\mu^-$ mass of 106 MeV/$c^2$ is converted to the nuclear excitation energy.
The produced nucleus deexcites by emitting neutrons, charged particles, and $\gamma$-rays, which results in the radionuclide production.

A considerable number of radionuclides are produced by the intense $\mu^-$ beam irradiation in the accelerator facilities.
The recent beam intensity upgrade in the accelerator facilities promotes several applications, such as non-destructive elemental analysis~\cite{Umegaki_2020, Biswas2022-tb, Nakamura_2023} and muon spin rotation and relaxation ($\mu^-$SR)~\cite{Sugiyama_2018, Sugiyama_2020, Umegaki_2022}.
In the applications, various samples are irradiated by the intense $\mu^-$ beam.
The intense $\mu^-$ beam irradiation produces a number of radionuclides in the irradiated samples.
{\color{black}For example, magnesium and titanium sample irradiation in the $\mu^-$SR experiment produced main radionuclides of $^{24, 25, 26}$Na as well as $^{23}$Ne and $^{44, 47, 48}$Sc together with $^{47}$Ca, respectively.
The produced radionuclides gave rise to activity of several tens of kBq}.

Production of radionuclides with long half-lives is of particular concern for radiation safety at the high-intensity accelerator facilities because of considerable sample activation by the intense $\mu^-$ beam.
The considerable sample radioactivity should be estimated for proper handling of the activated sample.
Estimation of the sample radioactivity using reliable radionuclide production probabilities is crucial for ensuring the safety of experiments at the accelerator facilities providing the intense $\mu^-$ beam.

Materials and Life Science Experimental Facility (MLF) of Japan Proton Accelerator Research Complex (J-PARC)~\cite{jparc_1999} provides intense pulsed muon and neutron beams from Muon Science Establishment (MUSE)~\cite{Higemoto2017-qp} and Japan Spallation Neutron Source (JSNS)~\cite{IKEDA20091}, respectively, using the 1-MW pulsed proton beam.
At the MLF, the radioactivity induced by $\mu^-$ irradiation can now be estimated with the updated SAmple Radioactivity Evaluation program (SARE-MLF).
The updated program has been created for calculating the $\mu^-$-induced radioactivity by expanding the former program developed for neutron-induced radioactivity.
In the update, a dataset of radionuclide production from $\mu^-$ nuclear capture~\cite{Yamaguchi_2024} has been incorporated into the former program.
The dataset of radionuclide production is based on the calculation by a Monte Carlo simulation code of Particle and Heavy Ion Transport code System (PHITS)~\cite{Sato2024-xu}, which includes muon interactions~\cite{Abe2017-xw}.
For practical use of the dataset in the updated SARE-MLF, validation of the dataset is essential and requires experimental radionuclide production probabilities on light-to-heavy nuclei.

Experimental radionuclide production probabilities have been reported by many research groups~\cite{Backenstoss1971-qj, Measday2006-ti, Measday2007-ed, Measday_2007_Al, Heisinger2002-dh, Heusser1972-oa, Wyttenbach1978-ss}.
Backenstoss \textit{et al.}~\cite{Backenstoss1971-qj} and Measday \textit{et al.}~\cite{Measday2006-ti, Measday2007-ed, Measday_2007_Al} obtained the probabilities on light-to-heavy targets by detecting prompt $\gamma$-rays from the excited nucleus produced in $\mu^-$ nuclear capture.
In the prompt $\gamma$-ray measurement, the number of produced nuclides can be extracted from the sum of intensities for the $\gamma$-rays to the ground state.
However, a part of the transitions should be missed due to weak transitions, unidentified $\gamma$-ray energies, and particle emissions to the ground state of the produced nuclide.
Thus, the prompt $\gamma$-ray measurement always gives a lower limit of the production probabilities.

On the other hand, activation experiments were adopted by the other groups~\cite{Heisinger2002-dh, Heusser1972-oa, Wyttenbach1978-ss}, which are more reliable in measuring radionuclide production probabilities.
Heisinger \textit{et al.} obtained the probabilities on ten target elements from oxygen ($Z = 8$) to copper ($Z = 29$)~\cite{Heisinger2002-dh}.
The probabilities have been presented for produced radionuclides with half-lives ten minutes to several years together with the previous measurement by Heusser \textit{et al.}~\cite{Heusser1972-oa}.
However, there are conflicts between the two experiments, for example, $^{24}$Na production in $\mu^-$ nuclear capture on the aluminum target.
In addition, the reported experimental data is insufficient for validating the dataset obtained by PHITS since the target $Z$ is limited to $Z \le 29$.
Although the experimental data with higher-$Z$ target have been reported by Wyttenbach \textit{et al.}~\cite{Wyttenbach1978-ss}, the presented probabilities are limited to the radionuclides produced by reactions involving charged particle emission.
These insufficient coverage and above conflicts between existing measured data imply that new experimental data should be measured for updating experimental data and validating the calculated dataset.

In this study, we have measured radionuclide production probabilities in $\mu^-$ nuclear capture to update experimental data and to validate the calculated dataset for low-to-high $Z$ targets by the activation experiment.
{\color{black}The targets of $^{27}$Al, $^\mathrm{nat}$Si, $^{59}$Co, and $^\mathrm{nat}$Ta are preferential choices from monoisotopic elements for simplifying discussion in the validation and from experimental sample elements at the MLF for practical use}.
By comparing the resultant probabilities with previous activation experiments and the calculated dataset, we have investigated the conflict between experimental probabilities and validity of the dataset for the radioactivity estimation, respectively.

\section{Experiment}
\label{sec:exp}

\begin{figure}[tpbh]
\centering
\includegraphics[width=3.4in,clip]{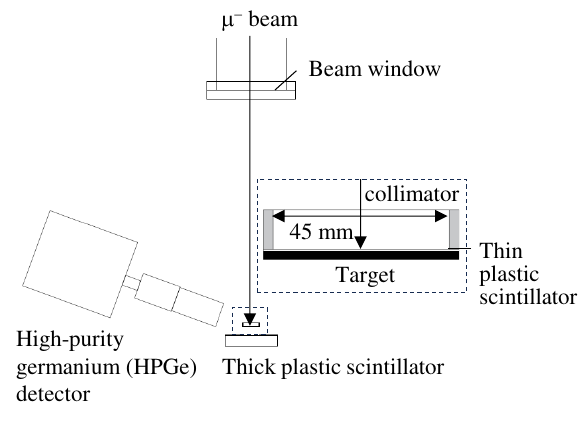}
	\caption{Plan view of experimental setup at D2 area. Incident negative muons come from the upper side of the figure through a beam window and stop in a target after passing through a beam collimator and a thin plastic scintillator. The components around the target are shown in an enlarged view.}
	\label{Setup}
\end{figure}

The experiment was performed using the pulsed $\mu^-$ beam from MUSE in the MLF, J-PARC.
The $\mu^-$ beam was produced by decays of negative pions from nuclear reactions between a 20-mm-thick graphite target and a pulsed 3-GeV proton beam at a repetition rate of 25 Hz.
Because the pulsed proton beam had two bunches with a 600-ns time interval, the pulsed $\mu^-$ beam had the same time structure.
The produced $\mu^-$ beam was transported to the D2 experimental area through the decay muon beamline (D-line) of MUSE.
Figure ~\ref{Setup} shows a plan view of the experimental setup used at the D2 area.
The setup in Fig.~\ref{Setup} comprises counting scintillators, a target, and a germanium $\gamma$-ray detector.

The incident $\mu^-$ beam from the D-line was extracted through a Kapton beam window with a diameter of 110 mm and a thickness of 0.1 mm.
The extracted beam was stopped in a target after passing through a beam collimator and a thin plastic scintillator placed in contact with the target.
The beam collimator was an acrylic plate of size 70 mm$\times50$\,mm and thickness 10 mm with a 45-mm square central aperture, which defined the beam irradiation region.
The thin plastic scintillator worked as a $\mu^-$ counter with an active area equal to the size of the collimator and a thickness of 0.5 mm.
A 20-mm-thick plastic scintillator {\color{black}masked by a tin plate with a 50-mm square central aperture} was placed downstream of the target and was used for selecting the momentum of the $\mu^-$ beam from the D-line by detecting the $\mu^-$ penetrating the target.
The beam momentum was selected to ensure that the $\mu^-$ penetrated the thin plastic scintillator and stopped in the target.
The selected momentum was 32\,MeV/c for the $\mu^-$ beam with a typical intensity of $1\times10^4$\,s$^{-1}$.

The targets used in the present experiment are listed in Table~\ref{target}.
All targets were 50-mm square plates with thickness between 0.5 and 0.9\,g/cm$^2$.
The elemental purity of each target was greater than 99.9\%.
Monoisotopic elements were preferentially selected for simplifying discussion on systematic trends of the radionuclide production probabilities by identifying the capturing nucleus without ambiguity.
Note that the natural abundance of $^{181}$Ta in the $^\mathrm{nat}$Ta target is greater than 99.9\%~\cite{natural_abundance}.
The $^\mathrm{nat}$Si target was also selected from experimental sample elements at the MLF for estimating silicon sample radioactivity based on the measurement.

\begin{table}[tpbh]
\centering
\caption{Thickness and elemental purity of targets and measurement time for in-beam and off-line measurements. The in-beam measurement time is almost equal to the irradiation time.}\label{target}
\begin{tabular}{c c c c c}
\hline
  Target            & Thickness  & Purity    & \multicolumn{2}{c}{Measurement time [h]} \\ \cline{4-5}
                    & [g/cm$^2$] & [\%]      & in-beam    & off-line \\\hline
  $^{27}$Al         & 0.54       & 99.999    & 3          & 18 \\
  $^\mathrm{nat}$Si & 0.58       & $>$99.9   & 3          & 3 \\
  $^{59}$Co         & 0.89       & 99.9      & 18         & 6.5 \\
  $^\mathrm{nat}$Ta & 0.83       & 99.95     & 16         & 18 \\ \hline
\end{tabular}
\end{table}

The radionuclides produced in the irradiated target were identified and counted using both in-beam and off-line activation methods for covering a wide range of half-lives.
The in-beam activation method provides production probabilities of radionuclides with short half-lives ranging from tens of milliseconds to several hours by detecting delayed $\gamma$-rays during intervals between beam pulses~\cite{Niikura2024-bz}.
For the in-beam measurement, a n-type high-purity germanium (HPGe) detector (ORTEC GMX20P4-70) was located 116\,mm from the target at the D2 area.
The GMX detector had a 0.5-mm-thick beryllium window and a coaxial HPGe crystal with a diameter of 53\,mm and a length of 51.4\,mm.
In the off-line measurement, a p-type HPGe detector (ORTEC GEM20P4-70) placed outside the D2 area was used for measuring the production probabilities of radionuclides with long half-lives.
The GEM detector had a 1-mm-thick aluminum window and a coaxial HPGe crystal with a diameter of 53.8\,mm and a length of 52.5\,mm.
In the off-line setup, a multi-layered shield surrounded the GEM detector for reducing background radiation. 
The shield consisted of a 10-cm-thick low-background lead and a graded liner of 1-mm-thick tin and 1.6-mm-thick copper to absorb X-rays from the lead.
For both GMX and GEM detectors, energy and efficiency were calibrated using $^{133}$Ba and $^{60}$Co standard $\gamma$-ray sources.

\begin{figure*}[ht]
\centering
\includegraphics[width=7.2in,clip]{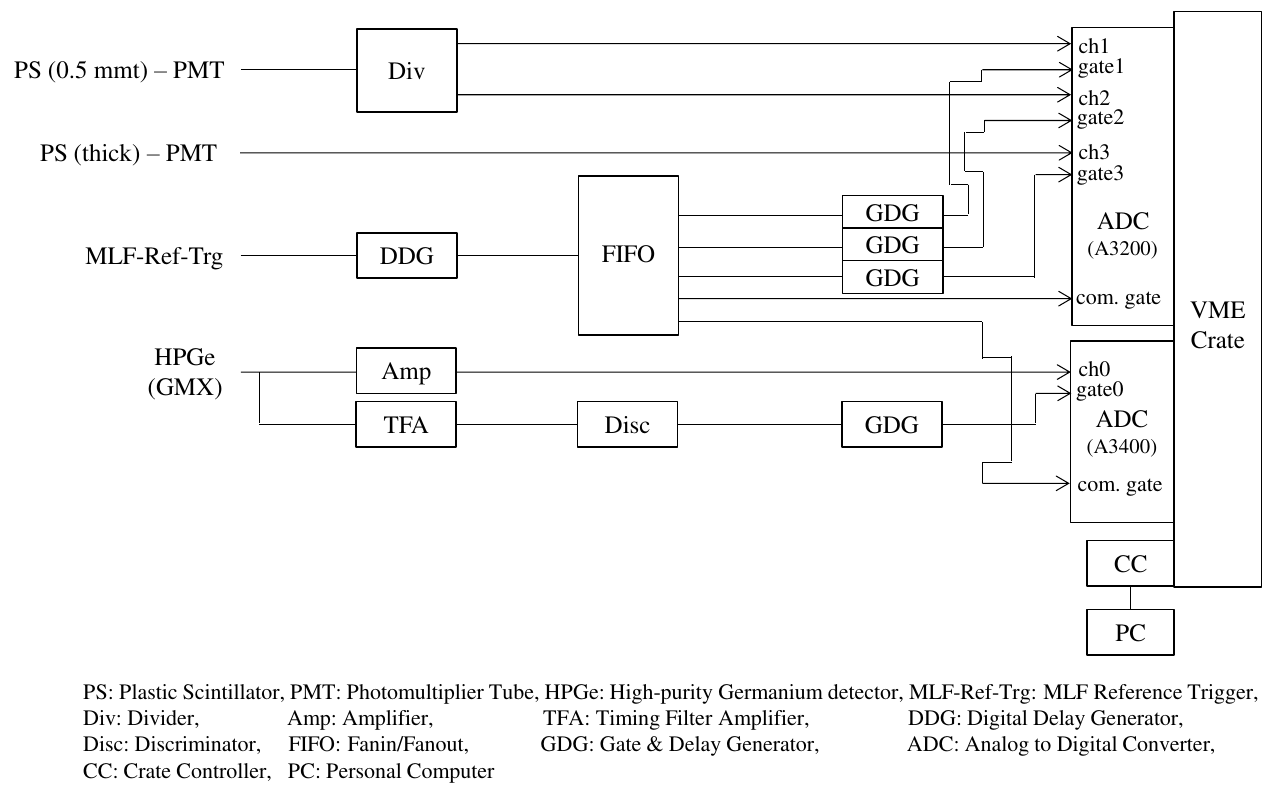}
	\caption{Block diagram of electronics for in-beam measurement. Modules of the electronics and the GMX detector are shared for D-line experiments. The MLF reference trigger is provided at 25 Hz based on the master clock of the J-PARC timing system.}
	\label{Circuit}
\end{figure*}
\begin{figure}
\centering
\includegraphics[angle=270,width=3.4in,clip]{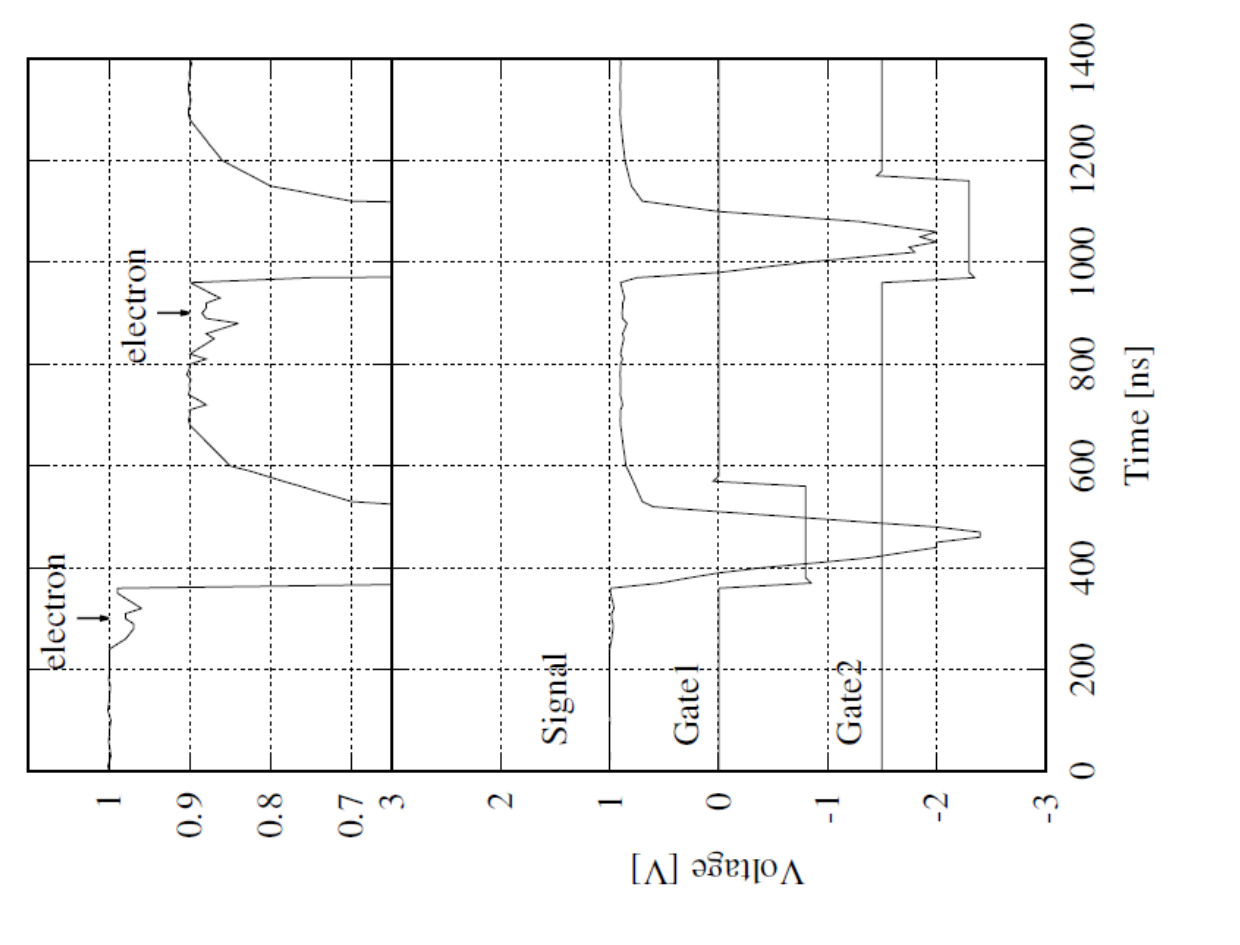}
	\caption{Waveform of thin plastic scintillator signal and gates for each bunch. The top figure is an enlarged view around the signal baseline.}
	\label{Waveform}
\end{figure}

The signals of the scintillation counters and the GMX detector were processed by electronics following the block diagram shown in Fig.~\ref{Circuit}.
The charge signal from a photomultiplier tube (Hamamatsu H11934-100) coupled with the thin plastic scintillator (Eljen EJ-212) was integrated by a charge-integrating analog-to-digital converter (ADC, NIKI GLASS A3200) for obtaining the number of the $\mu^-$ stopped in the target.
The thin plastic scintillator signal shown in Fig.~\ref{Waveform} was divided into signals for ch1 and ch2 inputs of the charge-integrating ADC.
The signals for ch1 and ch2 were integrated separately for the first and the second bunch, respectively, using gate signals synchronized with the timing of each bunch for avoiding counting electrons contained in the $\mu^-$ beam.
The gate signals were generated from the MLF reference trigger, which is a 25-Hz trigger.
The trigger signal was delayed for the beam timing through the digital delay generator (Stanford Research Systems DG645) and triggered individual gate signals of gate1 and gate2, which had the time interval of 600\,ns and the same width of approximately 200\,ns.
The signal of the GMX detector was fed into a peak sensing ADC (NIKI GLASS A3400) through an amplifier (ORTEC 572A) for obtaining energy spectra.
The gate signal for the peak sensing ADC was triggered by the GMX signal.
The ADCs of A3200 and A3400 were operated in triggered list mode, where timestamps and event counts were recorded, as well as ADC data of the integrated charge and the peak of the detector signals.
In this mode, the timestamp was initialized by a common gate signal, namely the 25-Hz trigger signal delayed for the beam timing, and gives the elapsed time from the common gate.
The event count was incremented by the common gate and tells us the measurement time.
The measurement time, which was almost equal to the irradiation time, is summarized in Table~\ref{target}. For the GEM detector in the off-line setup, the signal was fed into a multichannel analyzer (SEIKO EG\&G MCA-7) for obtaining energy spectra.


\section{Analysis}
\label{sec:ana}

The procedure of data analysis is described in detail in Ref.~\cite{Niikura2024-bz}, and important items are explained here.
When a $\gamma$-ray peak was observed for a produced radionuclide, the radionuclide production probability $P$ was obtained using
\begin{equation}
P=\frac{N_\gamma/(I_\gamma\varepsilon_\gamma\varepsilon_\mathrm{LT})}{P_\mathrm{cap}P_\mathrm{d}N_\mu},
\label{eq_prob}
\end{equation}
where $N_\gamma$ is the number of detected $\gamma$-rays of interest, $I_\gamma$ is the $\gamma$-ray intensity per decay of the parent nucleus, $\varepsilon_\gamma$ is the absolute photopeak efficiency of the HPGe detector, $\varepsilon_\mathrm{LT}$ is a factor for dead-time correction, namely analysis live-time, $P_\mathrm{cap}$ is the nuclear capture probability, $P_\mathrm{d}$ is the ratio of parent nucleus which decays during each measurement of in-beam and off-line to the total of the produced parent nucleus, and $N_\mu$ is the number of the $\mu^-$ stopped in the target.
In the case where multiple $\gamma$-rays were observed from a produced radionuclide, a weighted average was taken with the production probabilities obtained from each $\gamma$-ray using Eq.~(\ref{eq_prob}).

\begin{figure} 
\centering
\includegraphics[angle=0,width=3.4in,clip]{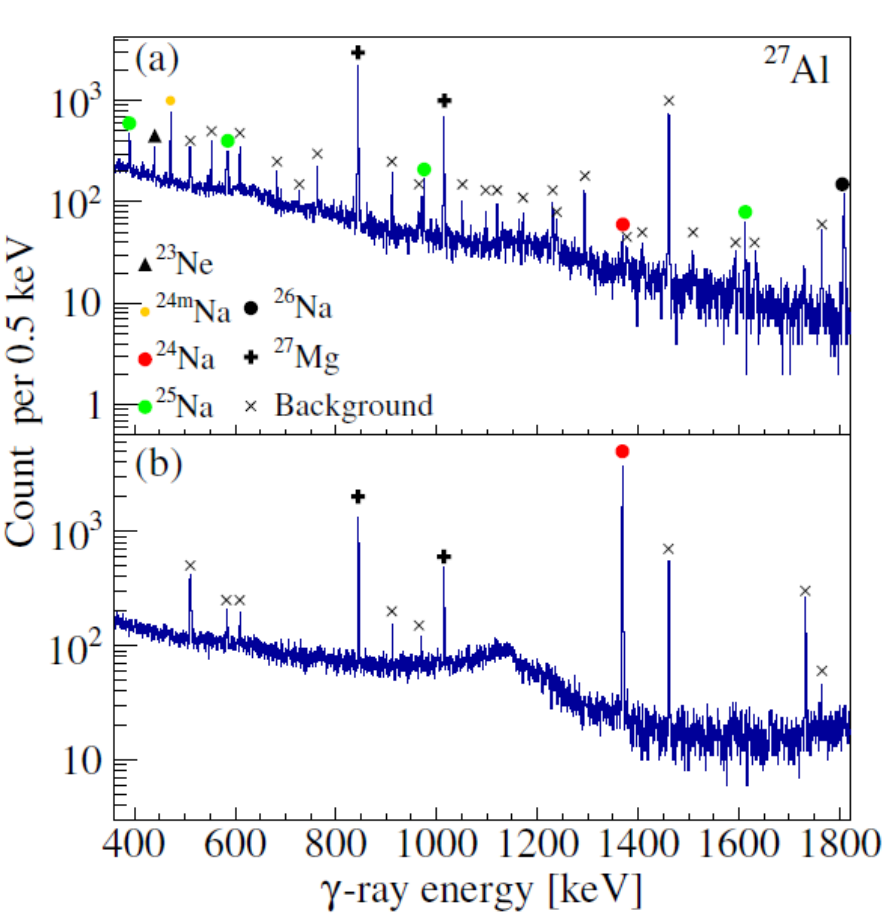}
	\caption{Energy spectra for $^{27}$Al target obtained by (a) in-beam and (b) off-line measurements. Peaks of $\beta$- or isomeric-decay $\gamma$-rays are marked with closed symbols.}
	\label{Alspectra}
\end{figure}

\begin{figure}
\centering 
\includegraphics[angle=0,width=3.4in,clip]{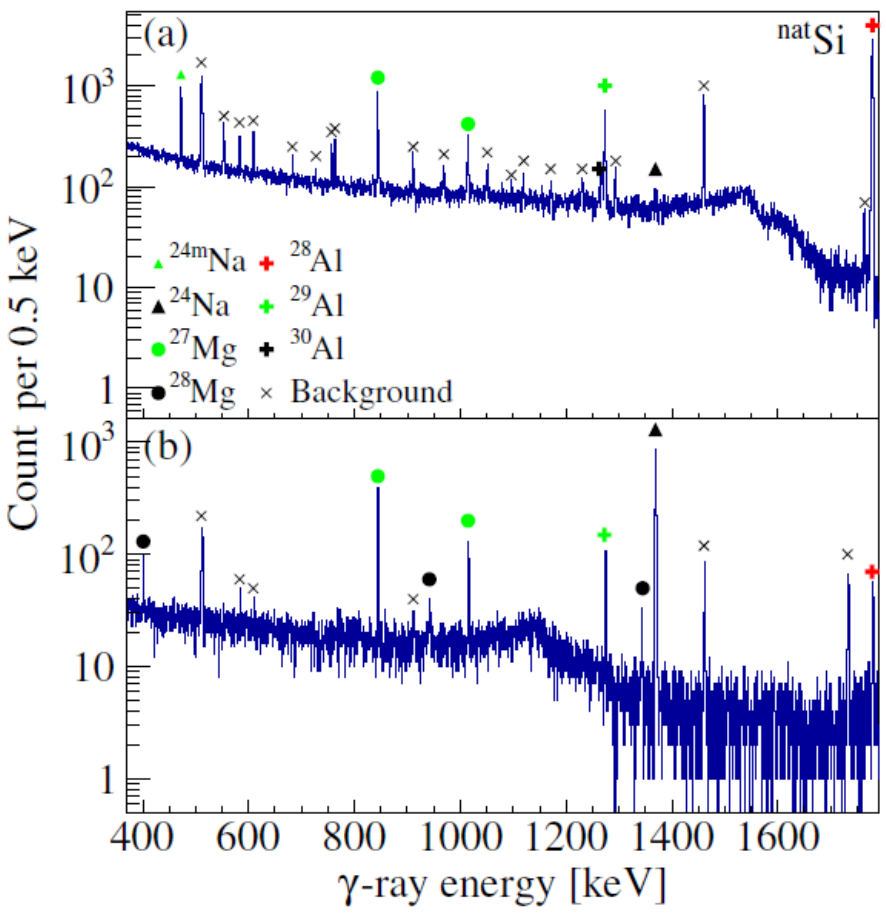}
	\caption{Energy spectra for $^\mathrm{nat}$Si target. (a) In-beam spectrum; (b) off-line spectrum. Peaks of $\beta$- or isomeric-decay $\gamma$-rays are marked with closed symbols.}
	\label{Sispectra}
\end{figure}

\begin{figure}
\centering 
\includegraphics[angle=0,width=3.4in,clip]{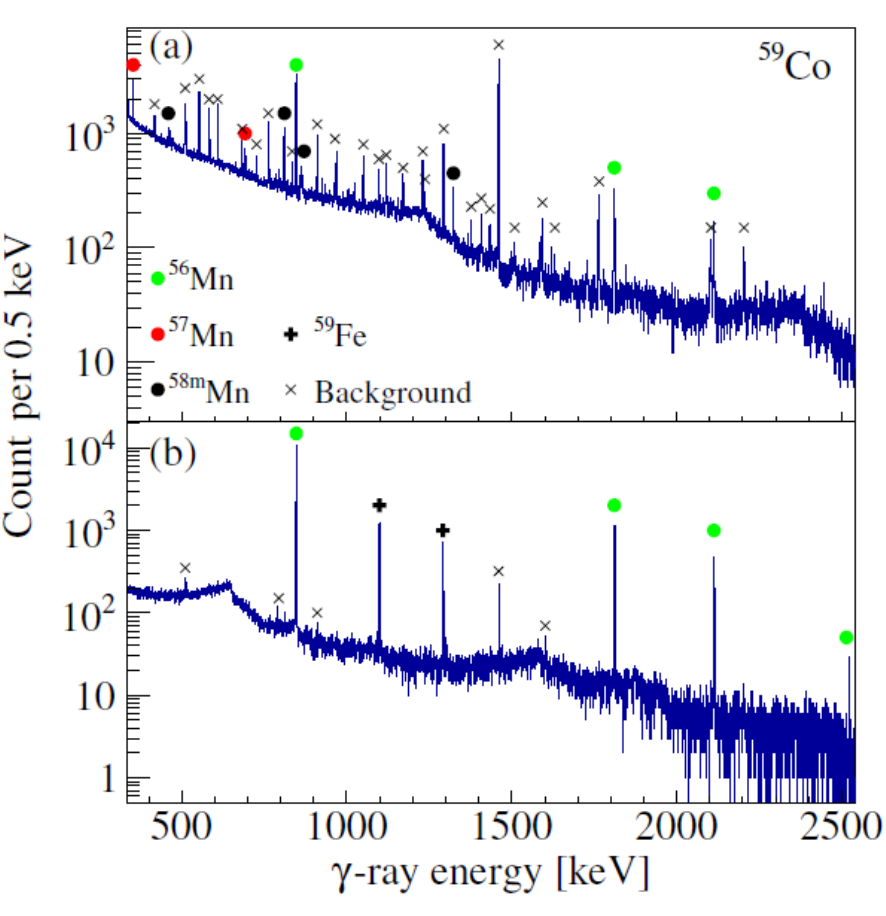}
	\caption{Energy spectra for $^{59}$Co target. (a) In-beam spectrum; (b) off-line spectrum. Peaks of $\beta$- or isomeric-decay $\gamma$-rays are marked with closed symbols.}
	\label{Cospectra}
\end{figure}

\begin{figure}
\centering 
\includegraphics[angle=0,width=3.4in,clip]{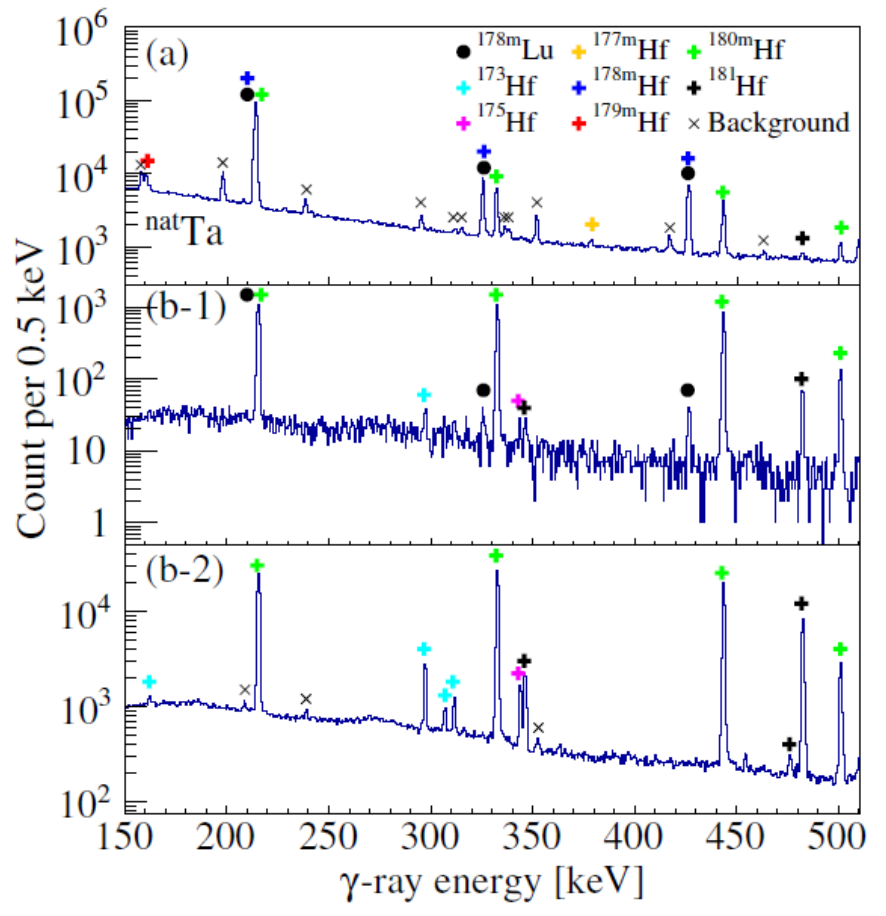}
	\caption{Energy spectra for $^\mathrm{nat}$Ta target. (a) In-beam spectrum; (b--1) off-line spectrum in the first ten minutes; (b--2) off-line spectrum over the entire measurement time. Peaks of $\beta$- or isomeric-decay $\gamma$-rays are marked with closed symbols.}
	\label{Taspectra}
\end{figure}

The number of $\gamma$-rays $N_\gamma$ was obtained by fitting a peak of interest in the measured energy spectra using a Gaussian function with background term.
Energy spectra for the $^{27}$Al, $^\mathrm{nat}$Si, $^{59}$Co, and $^\mathrm{nat}$Ta targets are shown in Figs.~\ref{Alspectra}--\ref{Taspectra}.
In Figs.~\ref{Alspectra}--\ref{Taspectra}, $\gamma$-ray peaks from radionuclides produced by $\mu^-$ nuclear capture are clearly observed.
{\color{black}Note that the background peaks originate from natural radionuclides and indium isotopes produced by $\mu^-$ nuclear capture in the tin mask}.

The $\gamma$-ray intensity $I_\gamma$ was taken from Evaluated Nuclear Structure Data File (ENSDF)~\cite{BASUNIA20211, BASUNIA20223, FIRESTONE20091691, BASUNIA20161, BASUNIA20111875, JUNDE20111513, BHAT1998415, NESARAJA2010897, BASUNIA20181, SHIRLEY1995377, BASUNIA2004719, ACHTERBERG20091473, BAGLIN2009265, MCCUTCHAN2015151, WU2005367}.
ENSDF typically provides relative intensities to the most intense $\gamma$-ray and a factor $f_\mathrm{abs}$ to convert the relative values to absolute intensities, including their uncertainties, although the uncertainties of relative intensities are sometimes missing for the most intense $\gamma$-ray.
The uncertainties of relative intensities were assigned to weights with statistical uncertainties of $N_\gamma$ for the averaging procedure of $P$.
In the case of the missing uncertainty, it was estimated from uncertainties of other $\gamma$-rays, assuming that the uncertainties are proportional to the square root of the relative intensities.
The uncertainty of the factor was taken into account after taking the weighted average.

The efficiency $\varepsilon_\gamma$ was measured using $^{133}$Ba and $^{60}$Co standard $\gamma$-ray sources which were set at the target position and had activities weaker than 5\,kBq.
A systematic uncertainty of 3\% was assigned to $\varepsilon_\gamma$ due to uncertainty of the source activities provided by the manufacturers.

The analysis live-time $\varepsilon_\mathrm{LT}$ was taken into account for the in-beam measurement because of the dead-time $T_\mathrm{d}$ caused by muonic X-rays {\color{black}as well as prompt $\gamma$-rays} entering the GMX close to the target.
With $T_\mathrm{d}$, $\varepsilon_\mathrm{LT}$ is expressed by
\begin{equation}
    \varepsilon_\mathrm{LT} = \frac{\int_{T_\mathrm{d}}^{T_\mathrm{p}}\mathrm{exp}(-\lambda t)\mathrm{d}t}{\int_0^{T_\mathrm{p}}\mathrm{exp}(-\lambda t)\mathrm{d}t},
\end{equation}
where $T_\mathrm{p}$ is the period of the pulse beam 40\,ms, $\lambda$ is the decay constant of produced radionuclides.
Because the radionuclides of interest have half-lives $T_{1/2} = \mathrm{ln}2/\lambda$ much longer than the pulse beam period, $T_{1/2} \gg 40$\,ms, $\varepsilon_\mathrm{LT}$ was approximated as
\begin{equation}
    \varepsilon_\mathrm{LT} = \frac{T_\mathrm{p}-T_\mathrm{d}}{T_\mathrm{p}}.
\label{eq_anallive}
\end{equation}
In the present experiment, the dead-time was corrected using Eq.~(\ref{eq_anallive}) with $T_\mathrm{d} = 0.5$ ms.
For the off-line measurement, the live-time was almost unity because of a low count rate.

The nuclear capture probability $P_\mathrm{cap}$ was deduced from the following equation:
\begin{equation}
    P_\mathrm{cap} = \frac{\Lambda_\mathrm{c}}{Q/\tau_{\mu^+}+\Lambda_\mathrm{c}},
\label{eq_Pcap}
\end{equation}
where $\Lambda_\mathrm{C}$ is the total nuclear capture rate, $Q$ is the Huff factor, and $\tau_{\mu^+}$ is the positive muon lifetime, that is 2.1969811(22)\,$\mu$s~\cite{PDG2024}. 
The values of $\Lambda_\mathrm{C}$ with uncertainty and $Q$ were taken from Ref.~\cite{Suzuki1987-aq} for each target element.
From Eq.~(\ref{eq_Pcap}), $P_\mathrm{cap}$ was calculated to be 0.6095(13), 0.6586(16), 0.9179(73), and 0.97038(24) for Al, Si, Co, and Ta, respectively.
The uncertainty of $\Lambda_\mathrm{C}$ largely results in the systematic uncertainty of $P_\mathrm{cap}$.

The decay ratio to the total production of the parent nucleus during the measurement time $P_\mathrm{d}$ was calculated based on the following equation:
\begin{equation}
    P_\mathrm{d} = \frac{\int\lambda n(t)\mathrm{d}t}{N_\mu},
\label{eq_pd}
\end{equation}
where $n(t)$ is the number of a produced radionuclide at time $t$ and was deduced assuming that the $\mu^-$ stopped in the target produces the radionuclide with $P = 1$. 
In the present analysis, discrete expressions were used because $P_\mathrm{d}$ can be calculated by the measured integrated charge of the thin plastic scintillator for each $\mu^-$ beam pulse.
For the in-beam measurement, the expression of Eq.~(\ref{eq_pd}) was converted into the discrete expression by time-averaging of $n(t)$ over $T_\mathrm{p}$ between pulses as follows:
\begin{equation}
    P_\mathrm{d} = \frac{\sum_i n_i\{1-\mathrm{exp}(-\lambda T_\mathrm{p})\}}{N_\mu},
\end{equation}
where the index $i$ is the event count, $n_i$ is the number of the radionuclide just after the beam pulse at the $i$-th event, and the summation range is the total event count.
For the off-line measurement,
\begin{equation}
    P_\mathrm{d} = \frac{\int_{t_{\mathrm{start}}}^{t_{\mathrm{stop}}}n_{i_\mathrm{tot}}\lambda\mathrm{exp}(-\lambda t)\mathrm{d}t}{N_\mu},
\label{eq_pd_offline}
\end{equation}
where $t_\mathrm{start}$ and $t_\mathrm{stop}$ are the start and stop timings of the measurement, respectively, and $n_{i_\mathrm{tot}}$ is the number of the radionuclide at the irradiation stop time.
In Eq.~(\ref{eq_pd_offline}), the irradiation stop time is defined as $t=0$.
The uncertainty of $P_\mathrm{d}$ arises from that of $T_{1/2}$ and is generally smaller than 0.1\%.

The number of the stopped $\mu^-$ $N_\mu$ was obtained as 
\begin{equation}
   N_\mu=f_\mu\sum_i q_i,
\end{equation}
where $f_\mu$ is a factor to convert the measured integrated charge into the absolute value and $q_i$ is the integrated charge for the beam pulse at the $i$-th event.
The conversion factor was deduced based on Eq.~(\ref{eq_prob}) for the 843.8-keV $\gamma$-ray from $^{27}$Mg production in the $^{27}$Al target using the known probability $P=$ 9.9(5)\%~\cite{Mizuno2025_PRC}.
The uncertainty of the reference value results in the uncertainty of the conversion factor, namely, the systematic uncertainty of $N_\mu$. 

\section{Results}
\label{sec:result}

The radionuclide production probabilities per $\mu^-$ nuclear capture are summarized in Table~\ref{result_p} with statistical and systematic uncertainties.
{\color{black}The systematic uncertainty includes the uncertainty of $f_\mathrm{abs}$, $\varepsilon_\gamma$, $P_\mathrm{cap}$, $P_\mathrm{d}$, and $f_\mu$}.
Table~\ref{result_p} provides the ground- and isomeric-state production probabilities separately.
It should be noted that the present results are cumulative production probabilities in many cases.
In particular, the $^{24}$Na production probability corresponds to the total probabilities of the directly produced $^{24}$Na and $^\mathrm{24m}$Na, which is an isomeric transition state (1$^+$, $T_{1/2} = 20$\,ms) and decays into the ground state.

The $^{58\rm m}$Mn and $^{178\rm m}$Lu isomeric states are $\beta$-decaying states and have the corresponding $\beta$-decaying ground states of $^{58}$Mn and $^{178}$Lu, respectively.
While the probabilities for the isomeric states were obtained, those for the ground states were not obtained because the $\gamma$-rays from the $\beta$-decays of $^{58}$Mn and $^{178}$Lu were not observed in the present measurement.
This may be due to low $I_\gamma$ for the $\gamma$-rays from the $\beta$-decays of $^{58}$Mn and $^{178}$Lu.
The $^{178\rm m}$Hf, $^{179\rm m}$Hf, and $^{180\rm m}$Hf isomeric transition states have the corresponding ground states of $^{178}$Hf, $^{179}$Hf, and $^{180}$Hf, respectively.
The ground states of $^{178}$Hf, $^{179}$Hf, and $^{180}$Hf were not observed because they are stable.
Although $^{178\rm m}$Hf and $^{179\rm m}$Hf have another isomeric transition state with a high spin of (16$^+$, $T_{1/2} = 31$\,y) and (25/2$^-$, $T_{1/2} = 25.05$\,d), respectively, these states were not observed.

It should be emphasized here that the production probability of $^{173}$Hf can be partially contributed by nuclear capture of negative pion ($\pi^-$), which can be contained in the $\mu^-$ beam because an average multiplicity of emitted neutron, 7.9 has been reported for $\pi^-$ nuclear capture on $^{181}$Ta~\cite{Beetz_1978}.
The $\pi^-$ contamination cannot be determined because no $\gamma$-rays from $\pi^-$ nuclear capture were observed except for $^{173}$Hf.
However, an upper limit of the $\pi^-$ contamination can be estimated from that of the $^{170}$Hf production because $^{170}$Hf is produced by $\pi^-$ nuclear capture in the probability of 0.131(4)~\cite{Beetz_1978} and may not be produced by $\mu^-$ nuclear capture.
The upper limit of the number of produced $^{170}$Hf was estimated to be $1.7\times10^4$ assuming that the 621-keV $\gamma$-ray peak exists in the off-line spectrum.
Therefore, the estimated upper limit of the $\pi^-$ contamination is $2\times10^{-4}$ per stopped $\mu^-$, which leads to relative contribution $<5\%$ to the $P$ of $^{173}$Hf.
\begin{table*}
\centering
\caption{Radionuclide production probabilities of present and previous experiments by Heisinger \textit{et al.}~\cite{Heisinger2002-dh}, Heusser \textit{et al.}~\cite{Heusser1972-oa}, and Wyttenbach \textit{et al.}~\cite{Wyttenbach1978-ss}. The produced radionuclides with half-life $T_{1/2}$ are also shown for each target together with energy $E_\gamma$ and intensity $I_\gamma$ of $\gamma$-rays detected by the method of in-beam or off-line. {\color{black}The calculated probabilities by PHITS in the last column are compared with the present ones in discussion}.}\label{result_p}
\scalebox{0.9}{
\begin{tabular}{c c c c c c c c c c}
\hline
  Target           & Radionuclide  & $T_{1/2}$    & $E_\gamma$ [keV] & $I_\gamma$ [\%]$^\mathrm{a}$ & Method & $P$ [\%] (Present)$^\mathrm{b}$ & $P$ [\%] (Previous)$^\mathrm{c}$ & {\color{black}$P$ [\%] (PHITS)$^\mathrm{d}$}\\ \hline 
  $^{27}$Al        & $^{23}$Ne      & 37.25 s      & 440.3                   & 32.9(2)$^\mathrm{e}$ & in-beam & 0.73(6)(9) & 0.76(11)~\cite{Wyttenbach1978-ss} & {\color{black}0.760(5)}\\
                   &                   &            &                  & $f_\mathrm{abs}$ = 0.329(10) &  & & \\
                     & $^{24}$Na      & 14.956 h     & 1368.6                 & 99.994(2)$^\mathrm{f}$ & off-line & 1.93(2)(15)$^\mathrm{g}$ & 2.1(2)~\cite{Heisinger2002-dh} & {\color{black}1.115(6)}\\
                     &                   &                 &                      &                 &           & & 3.5(8)~\cite{Heusser1972-oa} \\
                     & $^{25}$Na      & 59.1 s        & 389.7                   & 12.7(2)   & in-beam & 2.61(9)(37) & 2.8(4)~\cite{Wyttenbach1978-ss} & {\color{black}2.617(10)} \\
                     &                   &                 & 585.0               & 13.0(2)       &           & & \\
                     &                   &                 & 974.7               & 15.0(2)       &           & & \\
                     &                   &                 & 1611.7             & 9.48(14)      &           & & \\
                     &                   &                 &                      & $f_\mathrm{abs}$ = 0.130(7) &  & & \\
                     & $^{26}$Na      & 1.07128 s    & 1808.7             & 99.08(9)$^\mathrm{e}$     & in-beam & 0.84(6)(7) & & {\color{black}1.210(7)} \\
                     &                   &                 &                      & $f_\mathrm{abs}$ = 0.9908(3) &  & & \\
                     & $^{27}$Mg     & 9.458 min    & 843.8               & 71.80(2)$^\mathrm{f}$     & in-beam & 9.90(12)(79)$^\mathrm{h}$ & & {\color{black}15.85(2)}\\
                     &                   &                 & 1014.5             & 28.20(2)$^\mathrm{f}$     &           & & \\
  $^\mathrm{nat}$Si & $^{24}$Na     & 14.956 h      & 1368.6             & 99.994(2)$^\mathrm{f}$    & off-line & 1.71(4)(14)$^\mathrm{g}$ & \\
                     & $^{27}$Mg     & 9.458 min    & 843.8               & 71.80(2)$^\mathrm{f}$     & in-beam & 2.96(6)(24) & \\
                     &                   &                 & 1014.5             & 28.20(2)$^\mathrm{f}$     &            & & \\
                     & $^{28}$Mg     & 20.915 h      & 400.6               & 35.9$^\mathrm{i}$          & off-line & 0.12(1)(1) & \\
                     &                   &                 & 941.7               & 36.3$^\mathrm{i}$          &           & & \\
                     &                   &                 & 1342.2             & 54.0$^\mathrm{i}$          &           & & \\
                     & $^{28}$Al      & 2.245 min    & 1779.0              & 100$^\mathrm{i}$          & in-beam & 22.3(2)(18) & \\
                     & $^{29}$Al      & 6.56 min      & 1273.4              & 91.26(20)$^\mathrm{e}$   & in-beam & 2.75(6)(22) & \\
                     &                   &                 &                      & $f_\mathrm{abs}$ = 0.9126(6) &  & & \\
                     & $^{30}$Al      & 3.62 s         & 1263.1             & 41(1)        & in-beam & 0.32(7)(3) & \\
                     &                   &                 &                      & $f_\mathrm{abs}$ = 0.65(1) &  & & \\
  $^{59}$Co       & $^{56}$Mn     & 2.5789 h      & 846.8              & 98.9(7)$^\mathrm{e}$       & off-line  & 1.24(1)(11) & 1.10(15)~\cite{Wyttenbach1978-ss} & {\color{black}1.553(6)} \\
                  &                   &                 & 1810.7             & 26.9(4)       &           & & \\
                     &                   &                 & 2113.1             & 14.2(3)      &           & & \\
                     &                   &                 & 2523.1             & 1.02(2)      &           & & \\
                     &                   &                 &                      & $f_\mathrm{abs}$ = 0.9885(3) &  & & \\
                     & $^{57}$Mn      & 85.4 s        & 352.3               & 2.1(1)       & in-beam & 1.50(11)(70) & 2.4(3)~\cite{Wyttenbach1978-ss} & {\color{black}2.985(8)} \\
                     &                   &                 & 692.0               & 5.5(2)       &           & & \\
                     &                   &                 &                      & $f_\mathrm{abs}$ = 0.0055(21) &  & & \\
                     & $^\mathrm{58m}$Mn & 65.4 s     & 459.2               & 21(1)       & in-beam & 0.26(1)(4) & 0.19(2)~\cite{Wyttenbach1978-ss} \\
                     &                   &                 & 810.8               & 88(3)       &           & & \\
                     &                   &                 & 863.9               & 15(1)       &           & & \\ 
                     &                   &                 & 1323.1             & 59(2)        &           & & \\
                     &                   &                 &                      & $f_\mathrm{abs}$ = 0.88(4) &  & & \\
                     & $^{59}$Fe      & 44.490 d      & 1099.3             & 56.5(15)     & off-line & 11.9(3)(13) & 15.1(11)~\cite{Heusser1972-oa} & {\color{black}14.37(2)} \\
                     &                   &                 & 1291.6             & 43.2(11)     &            & & \\
                     &                   &                 &                      & $f_\mathrm{abs}$ = 0.565(11) &  & & \\
  $^\mathrm{nat}$Ta & $^\mathrm{178m}$Lu & 23.1 min & 325.6              & 94.1(11)     & off-line & 0.030(4)(4) & 0.04(1)~\cite{Wyttenbach1978-ss} & {\color{black}0.675(8)} \\
                    &                   &                 & 426.4               & 97.0(13)     &           & & \\
                     &                   &                 &                      & $f_\mathrm{abs}$ = 0.941(12) &  & & \\
                     & $^{173}$Hf    & 23.6 h         & 297.0               & 34(1)        & off-line & 0.11(1)(2)$^\mathrm{j}$ & \\
                     &                   &                 & 306.6               & 6.4(1)       &           & & \\
                     &                   &                 & 311.2               & 11(1)       &           & & \\
                     &                   &                 &                      & $f_\mathrm{abs}$ = 0.083(3) &  & & \\
                     & $^{175}$Hf     & 70 d          & 343.4              & 84(1)$^\mathrm{e}$        & off-line & 1.3(1)(2) & & {\color{black}0.0313(8)} \\
                     &                   &                 &                      & $f_\mathrm{abs}$ = 0.84(3) &  & & \\
                     & $^\mathrm{177m}$Hf & 1.09 s    & 378.5              & 38.03(36)    & in-beam & 0.0607(75)(61) & & {\color{black}0.358(31)} \\
                     &                   &                 &                      & $f_\mathrm{abs}$ = 0.1640(5) &  & & \\
                     & $^\mathrm{178m}$Hf & 4.0 s      & 325.6              & 94.1(11)    & in-beam & 0.85(1)(9) & & {\color{black}1.515(27)} \\ \hline                  
\end{tabular}
}
\end{table*}
\setcounter{table}{1}
\begin{table*}
\centering
\caption{(Continued.)}
\scalebox{0.9}{
\begin{tabular}{c c c c c c c c c}
\hline
  Target           & Radionuclide  & $T_{1/2}$    & $E_\gamma$ [keV] & $I_\gamma$ [\%]$^\mathrm{a}$ & Method & $P$ [\%] (Present)$^\mathrm{b}$ & $P$ [\%] (Previous)$^\mathrm{c}$ & {\color{black}$P$ [\%] (PHITS)$^\mathrm{d}$} \\ \hline 
                     &                   &                 & 426.4               & 97.0(13)    &           & \\ 
                     &                   &                 &                      & $f_\mathrm{abs}$ = 0.941(12) &  & \\
                     & $^\mathrm{179m}$Hf & 18.67 s   & 160.7              & 2.88(18)    & in-beam & 4.46(32)(40) & & {\color{black}7.98(32)}\\
                     &                   &                 &                      & $f_\mathrm{abs}$ = 0.953(1) &  & \\
                     & $^\mathrm{180m}$Hf & 5.53 h    & 332.3               & 94.1(24)    & in-beam & 1.05(2)(10) & & {\color{black}2.645(62)} \\
                     &                   &                 & 443.2               & 81.7(24)    &           & \\
                     &                   &                 & 500.7               & 14.2(4)      &           & \\
                     &                   &                 &                      & $f_\mathrm{abs}$ = 0.941(9) &  & \\
                     & $^{181}$Hf     & 42.39 d       & 345.9              & 15.1(1)       & off-line & 7.17(14)(72) & & {\color{black}9.650(14)} \\
                     &                   &                 &                      & $f_\mathrm{abs}$ = 0.805(4) & & \\ \hline
\end{tabular}
}
\begin{tablenotes}
\item[a]$^\mathrm{a}$Only uncertainty of relative intensity is taken into account unless noted.
\item[b]$^\mathrm{b}$Statistical and systematic uncertainties are written in parentheses in order.
\item[c]$^\mathrm{c}$Total uncertainty is written in the parenthesis.
\item[d]{\color{black}$^\mathrm{d}$Statistical error is written in the parenthesis}.
\item[e]$^\mathrm{e}$Uncertainty of relative intensity is not provided in ENSDF and is estimated from other uncertainties.
\item[f]$^\mathrm{f}$Only absolute intensity with uncertainty is provided in ENSDF.
\item[g]$^\mathrm{g}$Total probability of directly produced $^{24}$Na and $^\mathrm{24m}$Na.
\item[h]$^\mathrm{h}$Reference probability for $N_\mu$ calibration.
\item[i]$^\mathrm{i}$No uncertainty is provided in ENSDF.
\item[j]$^\mathrm{j}$Relative contribution from $\pi^-$ contamination is estimated to be less than 5\%.
\end{tablenotes}
\end{table*}

\section{Discussion}
\label{sec:disc}

We will first compare the present result with previous experiments.
As mentioned in Section~\ref{sec:intro}, Heisinger \textit{et al.}~\cite{Heisinger2002-dh} and Heusser \textit{et al.}~\cite{Heusser1972-oa} obtained the $^{24}$Na production probability in $\mu^-$ nuclear capture on $^{27}$Al.
The number of produced $^{24}$Na was measured by the common method to detect $\beta$-delayed $\gamma$-rays with a Ge detector of an off-line setup in the present and previous experiments.
The present $^{24}$Na production probability is 1.93(2)(15)\%, {\color{black}where the statistical and systematic uncertainties are in parentheses in order}, while the previous ones by Heisinger \textit{et al.} and Heusser \textit{et al.} are 2.1(2)\% and 3.5(8)\%, respectively.
The former agrees with the present result, while the latter is higher.
The production probabilities by Heusser \textit{et al.} on $^\mathrm{nat}$Fe, $^{59}$Co, and $^\mathrm{nat}$Ni targets are also higher than those by Heisinger \textit{et al.} and the present result.
The systematic discrepancy seems to arise from the method to obtain the number of the $\mu^-$ stopped in the target.
Heisinger \textit{et al.} measured the number of the stopped $\mu^-$ $N_\mu$ by detecting muonic X-rays.
The present experiment did by detecting the $\mu^-$ in each beam pulse with the thin plastic scintillator calibrated using the known production probability.
In contrast, Heusser \textit{et al.} derived $N_\mu$ from the total number of all produced nuclei in the $^\mathrm{nat}$Ni activation with a theoretical assumption.
Under the assumption, the production probability of the stable nucleus $^{59}$Co was estimated to be 15.9\% based on theoretical calculation.
In the derivation of $N_\mu$, stable iron nucleus production may be ignored despite the significant $^{56}$Fe production probability greater than 10\%~\cite{Heisinger2002-dh}.
The theoretical assumption and the stable iron treatment should result in the systematic discrepancy.

Another experiment by Wyttenbach \textit{et al.}~\cite{Wyttenbach1978-ss} provides production probabilities of radionuclides from charged particle emission.
The number of the produced radionuclides in $\mu^-$ nuclear capture was measured by the same method as the other three experiments.
The number of the $\mu^-$ stopped in the target was measured with a counter telescope consisting of four plastic scintillators.
The probabilities obtained by the measurement are compared with the present result in Table~\ref{result_p} and are reasonably consistent.

\begin{figure}
\centering
\includegraphics[angle=270,width=3.1in,clip]{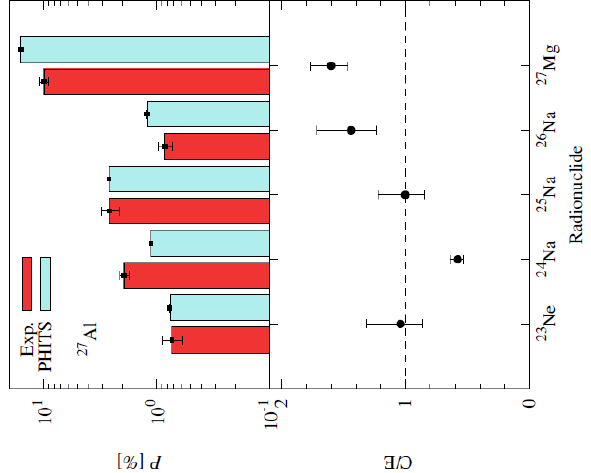}
	\caption{Radionuclide production probabilities and calculation/experiment (C/E) values on $^{27}$Al target. Calculations are PHITS results and experiments are present results.}
	\label{Alcomp}
\end{figure}

\begin{figure}
\centering 
\includegraphics[angle=270,width=3.1in,clip]{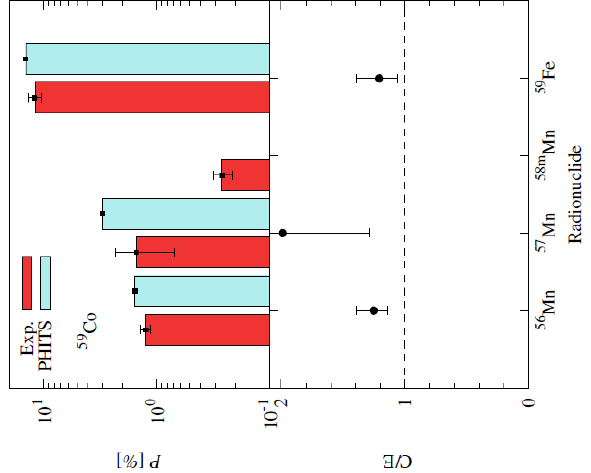}
	\caption{Radionuclide production probabilities and C/E values on $^{59}$Co target. The calculation and the experiment are the PHITS and present result, respectively. The probability of $^\mathrm{58m}$Mn is not obtained by PHITS calculation.}
	\label{Cocomp}
\end{figure}

\begin{figure}
\centering 
\includegraphics[angle=270,width=2.9in,clip]{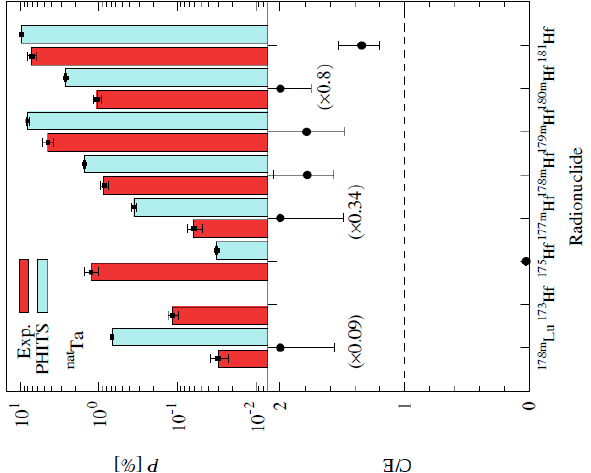}
	\caption{Radionuclide production probabilities and C/E values on $^\mathrm{nat}$Ta target. The calculation and the experiment are the PHITS and present result, respectively. The probability of $^{173}$Hf is not obtained by PHITS calculation. The large C/E values are plotted at $\mathrm{C/E}=2$ by multiplying the factor in the parenthesis.}
	\label{Tacomp}
\end{figure}

We will now compare the present results with the dataset calculated by PHITS.
In Figs.~\ref{Alcomp}-\ref{Tacomp}, present probabilities (red bar) are shown together with those from the dataset (cyan bar) for the $^{27}$Al, $^{59}$Co, and $^\mathrm{nat}$Ta targets, respectively, and are compared by calculation/experiment (C/E) values.
The dataset was obtained using PHITS version 3.20 with default input parameters in principle.
PHITS offers models for muon interactions, including $\mu^-$ nuclear capture, and describes it as follows~\cite{Abe2017-xw}.
The $\mu^-$ nuclear capture reaction gives the excitation energy to a neutron produced in the capturing nucleus.
The nuclear excitation energy is sampled from the distribution function proposed by Singer~\cite{Singer1962-ww} applying the momentum distribution of the proton in the nucleus by Amado~\cite{Amado1976-cy}.
Then, the time evolution of the nucleon system with the excitation energy is calculated using JAERI quantum molecular dynamics (JQMD) model~\cite{Niita1995-jqmd}.
After the time evolution is finished, particle emission in the evaporation process is described by generalized evaporation model (GEM)~\cite{Furihata_2001} and is followed by $\gamma$-ray emission.
The deexcitation by $\gamma$-ray emission is calculated using ENSDF-Based Isomeric Transition and isomer production Model (EBITEM)~\cite{Ogawa2014-jv}.

The calculated probabilities in Figs.~\ref{Alcomp}--\ref{Tacomp} generally overestimate present results by less than a factor of two.
The calculations for the $^{27}$Al target in Fig.~\ref{Alcomp} agree with the present results for $^{23}$Ne and $^{25}$Na production.
The calculated production probabilities of $^{26}$Na and $^{27}$Mg follow the general trend.
However, the calculated probability of $^{24}$Na mainly produced by $(\mu^-, \nu_\mu p2n)$ reaction underestimates the present result by a factor of 1.6.
For the $^{59}$Co target in Fig.~\ref{Cocomp}, the calculations of $^{56}$Mn, $^{57}$Mn, and $^{59}$Fe follow the general trend.
However, the production probability of the isomer $^{58\rm m}$Mn was not obtained by the calculation.
In Fig.\ref{Tacomp}, the calculations of $^{178\rm m}$Hf, $^{179\rm m}$Hf, and $^{181}$Hf for the $^\mathrm{nat}$Ta target follow the general trend.
However, the calculated probability of $^{178\rm m}$Lu mainly produced by the $(\mu^-, \nu_\mu p2n)$ reaction overestimates the present result by a factor of twenty.
In contrast, the calculated production probability of $^{175}$Hf from multiple neutron emission underestimates by a factor of fifty.
The production probability of $^{173}$Hf from eight-neutron emission was not obtained by the calculation.
The calculations of isomers $^{177\rm m}$Hf and $^{180\rm m}$Hf overestimates by more than a factor of two.

The overestimation and the absence of the isomer production probabilities are attributed to EBITEM.
EBITEM describes isomer production from the nuclear deexcitation by tracking $\gamma$-ray transitions for an isomeric state.
In the low energy states below 3 MeV, the deexcitation is calculated based on the level structure data in ENSDF.
Although the level structure data cover numerous levels, some nuclei have poor level structure data.
The insufficient data may result in the overestimation and the absence of calculated isomer production probabilities.

\begin{figure}
\centering
\includegraphics[angle=270,width=3.4in,clip]{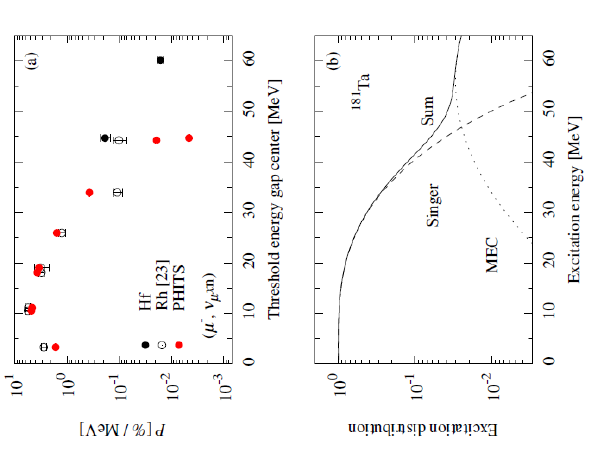}
	\caption{(a) Threshold energy dependence of measured production probabilities per threshold energy gap for Hf (black circles) and Rh~\cite{Niikura2024-bz} (open circles) isotopes on $^\mathrm{nat}$Ta and $^{106, 108}$Pd targets, respectively. Calculations by PHITS (red circles) are also displayed. (b) Excitation distribution for $^{181}$Ta. The function proposed by Singer is shown with the dashed line. The meson exchange current (MEC) contribution approximated by the Gaussian function is also shown with a dotted line. The solid line indicates the sum of dashed and dotted lines.}
	\label{Eth_Eex}
\end{figure}

The discrepancies in the production probability on the multiple neutron emission are investigated with respect to threshold energy dependence.
The present production probabilities of $^{173}$Hf and $^{175}$Hf are plotted in Fig.~\ref{Eth_Eex}(a) against the center of the threshold energy gap between $E_x$ and $E_{x+1}$ for $(\mu^-, \nu_\mu xn)$ together with those of rhodium isotopes on enriched palladium targets~\cite{Niikura2024-bz}.
The probabilities in Fig.~\ref{Eth_Eex}(a) indicate those per threshold energy gap.
Calculated production probabilities are also displayed in Fig.~\ref{Eth_Eex}(a) on the Hf and Rh isotopes for comparing the threshold energy dependence between measured and calculated results.
The threshold energies were calculated based on the mass table of AME2020~\cite{Wang2021-zy}.
Both measured and calculated results have a broad peak at 10 MeV and decrease exponentially from 10 to 45 MeV.
However, around 45 MeV, calculated probabilities more sharply decrease than measured ones.
The sharp decrease of calculation should result from the excitation distribution by Singer~\cite{Singer1962-ww} as shown with the dashed line in Fig.~\ref{Eth_Eex}(b).
The excitation distribution by Singer also shows the sharp decrease in the high excitation energy range.
This sharp decrease results in the underestimation and the absence of the calculated production probabilities of $^{175}$Hf and $^{173}$Hf from the multiple neutron emission, respectively.
The underestimation and the absence are expected to be improved by considering meson exchange current (MEC) approximated using Gaussian function~\cite{LIFSHITZ1988684,Minato2023-hr} shown with the dotted line in Fig.~\ref{Eth_Eex}(b) because of the MEC contribution in the high excitation energy range.

The underestimation and the overestimation in the calculated production probability on the $(\mu^-, \nu_\mu p2n)$ reaction are also investigated.
It should be noted that the $(\mu^-, \nu_\mu p2n)$ reaction cannot be distinguished from $(\mu^-, \nu_\mu dn)$ and $(\mu^-, \nu_\mu t)$ reactions by the activation experiment.
\begin{figure}
\centering
\includegraphics[angle=270,width=3.4in,clip]{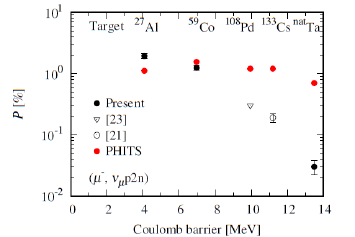}
	\caption{Radionuclide production probabilities on main reaction $(\mu^-, \nu_\mu p2n)$ as function of Coulomb barrier. Black circles: present data; open triangle: measured data (upper limit) by Niikura \textit{et al.}~\cite{Niikura2024-bz}; open circle: measured data by Wyttenbach \textit{et al.}~\cite{Wyttenbach1978-ss}; red circles: PHITS calculation.}
	\label{Coulomb}
\end{figure}
\begin{figure}
\centering
\includegraphics[angle=270,width=3.4in,clip]{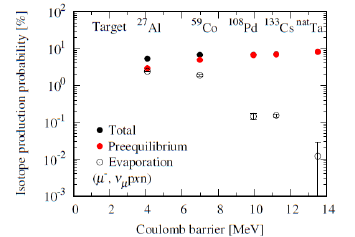}
	\caption{Calculated isotope production probabilities on $(\mu^-, \nu_\mu pxn)$ as function of Coulomb barrier. Although the Coulomb barrier is different among isotopes on a target, the value for the $(\mu^-, \nu_\mu p2n)$ reaction is used because of small difference. Black circles: total of the preequilibrium and evaporation contributions; red circles: contribution of preequilibrium process; open circles: contribution of evaporation process.}
	\label{Coulomb_pxn}
\end{figure}
However, experimental proton yield ratios to deuterons and tritons approximately greater than 3 have been reported~\cite{Budyashov_1971}.
Thus, $(\mu^-, \nu_\mu p2n)$ should be the main reaction.
Because the reaction involves charged particle emission, the present production probabilities should have Coulomb barrier dependence as reported by Wyttenbach \textit{et al.}~\cite{Wyttenbach1978-ss}.
In Fig.~\ref{Coulomb},  on Coulomb barrier dependence is compared between measured and calculated probabilities.
Coulomb barrier $V$ was calculated from
\begin{equation}
    V = \frac{e^2}{4\pi\varepsilon_0}\frac{Z_\mathrm{C}}{r_0A_\mathrm{C}^{1/3}},
\end{equation}
where $Z_\mathrm{C}$ and $A_\mathrm{C}$ are the atomic number and the mass number of the produced nuclide, respectively.
The constants of $e^2/(4\pi\varepsilon_0)$ and $r_0$ were set to 1.44 MeV and 1.35 fm, respectively.
The measured production probabilities in Fig.~\ref{Coulomb} show an exponential decrease with increasing the Coulomb barrier.
This trend is consistent with that reported by Wyttenbach \textit{et al.}.
In contrast, the calculated production probabilities are around 1\% and appear to be independent of the Coulomb barrier.
The $V$-independent behavior of the calculations should arise from the description of proton emission.
The proton emission in $\mu^-$ nuclear capture is described by the JQMD model as well as GEM.
The JQMD model describes energetic proton emission from the preequilibrium state, while GEM low energy proton emission in the evaporation process.
For identifying the poorly described process, each process was investigated on the proton emission $(\mu^-, \nu_\mu pxn)$ contributing to the $(Z-2)$ isotope production.
In Fig.~\ref{Coulomb_pxn}, the calculated isotope production probabilities are shown for each contributing process as well as the total.
The production probabilities from the evaporation process show the exponential decrease with increasing the Coulomb barrier.
This trend is consistent with that of the measured probabilities in Fig.~\ref{Coulomb}.
In contrast, the production probabilities from the preequilibrium process show a small increase.
Therefore, the preequilibrium process is poorly described by the JQMD model.
The increase trend of the preequilibrium process results in the $V$-independent behavior due to the large contribution and should lead to the discrepancy in particular 20-times overestimation on high-$Z$ targets.

From the point of view of radiation safety, the general trend of the calculation to overestimate by less than a factor of two is preferable to underestimation because of the {\color{black}safe} side.
However, the overestimations in the isomer production probabilities of $^{177\rm m}$Hf ($T_{1/2}=1.1$ s) and $^{180\rm m}$Hf ($T_{1/2}=5.5$ h) are relatively large and lead to an unreasonable estimation of sample radioactivity.
The 20-times overestimation for the $^{178\rm m}$Lu ($T_{1/2}=23$ min) production in the $(\mu^-, \nu_\mu p2n)$ reaction is too large.
In the $(\mu^-, \nu_\mu p2n)$ reaction, underestimation is also observed for the $^{24}$Na ($T_{1/2}=15$ h) production.
The 50-times underestimation for the $^{175}$Hf ($T_{1/2}=70$ d) production by the multiple neutron emission should be a problem in using the intense $\mu^-$ beam.
For correcting the underestimation in the multiple neutron emission, additional experimental probabilities of high threshold energy are necessary.
Further experimental data is also required for correcting the overestimation in the isomer production.
For the $(\mu^-, \nu_\mu p2n)$ reaction, the description of proton emission should be corrected by improving the JQMD model.

\section{Conclusion}
\label{sec:conc}
We measured radionuclide production probabilities in $\mu^-$ nuclear capture to update experimental data and to validate the dataset based on the PHITS calculation.
The probabilities were obtained for the $^{27}$Al, $^\mathrm{nat}$Si, $^{59}$Co, and $^\mathrm{nat}$Ta targets.
Present production probabilities compared with previous experiments resolved the conflict of $^{24}$Na production probability on the $^{27}$Al target and were confirmed to be reasonably consistent with previous experiments.
By comparing measured probabilities with the calculated dataset, we found that the calculated dataset is on the {\color{black}safe} side and is valid for estimating sample radioactivity except for the following three cases: (i) isomer production; (ii) radionuclide production by the multiple neutron emission with high threshold energy; (iii) radionuclide production by particle emissions involving a proton.
The case (iii) indicates that the calculation needs to be improved on Coulomb barrier dependence.
For (i) and (ii), further experimental data need to be taken for improving the calculation and for reasonable estimation of radioactivity.

\section*{Acknowledgments}
The authors are grateful to Shogo Doiuchi (KEK) for preparing the data acquisition system.
We would like to thank Hideaki Isozaki and Keiichi Inoue (Neutron Science Section, J-PARC) for their help in preparing targets and also thank Mitsuhiro Usami (Engineering Services Department, Japan Atomic Energy Agency) for machining the beam collimator.
We also thank members of Muon Science Section of J-PARC for their support in conducting the experiment.
The muon experiment at the Materials and Life Science Experimental Facility of J-PARC was performed under a user program (Proposal No. 2022B0185).



\bibliographystyle{elsarticle-num} 
\bibliography{paperpile_niikura,references}







\end{document}